\documentclass[preprint,12pt]{elsarticle}

\usepackage{graphicx}

\usepackage{amssymb}

\journal{Journal of Computational Physics}

\begin{document}

\begin{frontmatter}



\title{Full-wave parallel dispersive finite-difference time-domain modeling of three-dimensional electromagnetic cloaking structures}

\author{Yan Zhao}
\ead{yan.zhao@elec.qmul.ac.uk}
\author{and Yang Hao}

\address{School of Electronic Engineering and Computer Science, Queen Mary University of London, Mile End Road, London E1 4NS, United Kingdom}

\begin{abstract}
A parallel dispersive finite-difference time-domain (FDTD) method for the modeling of three-dimensional (3-D) electromagnetic cloaking structures is presented in this paper. The permittivity and permeability of the cloak are mapped to the Drude dispersion model and taken into account in FDTD simulations using an auxiliary differential equation (ADE) method. It is shown that the correction of numerical material parameters and the slow switching-on of source are necessary to ensure stable and convergent single-frequency simulations. Numerical results from wideband simulations demonstrate that waves passing through a three-dimensional cloak experience considerable delay comparing with the free space propagations, as well as pulse broadening and blue-shift effects.
\end{abstract}


\end{frontmatter}

\section{Introduction}
Recently, a great deal of attention has been paid to the analysis and design of electromagnetic cloaking structures, since first proposed by Pendry \textit{et al.} \cite{Pendry}. The specially designed cloak is able to guide waves to propagate around its central region, rendering the objects placed inside invisible to external electromagnetic radiations. The ideal cloak in \cite{Pendry} requires inhomogeneous and anisotropic media, with both permittivity and permeability independently controlled and radially dependent, making its practical realization very difficult. Therefore it has been proposed to use simplified material parameters for both transverse electric (TE) \cite{Cummer} and transverse magnetic (TM) \cite{Cai} cases. In order to reduce the scattering due to the impedance mismatch introduced by the simplified cloaks, an improved linear cloak \cite{YanOE}, a high-order transformation based cloak \cite{Cai2}, and a `square root' transformation based cloak \cite{ZhangOE} have also been proposed.

The coordinate transformation technique used in \cite{Pendry,Leonhardt} has also been applied to the design of magnifying perfect and super lenses \cite{Tsang}, electromagnetic field rotators \cite{ChenRotate}, the reflectionless complex media for shifting and splitting optical beams \cite{Rahm}, and conformal antennas \cite{Luo} etc. The experimental demonstration of a simplified cloak consisting of split-ring resonators (SRRs) has been reported at microwave frequencies \cite{Schurig}. For the optical frequency range, the cloak can be constructed by embedding silver wires in a dielectric medium \cite{Cai}, or using a gold-dielectric concentric layered structure \cite{Huang,Smolyaninov}.

The modeling of Pendry's invisible cloak has been performed by using both analytical and numerical methods. Besides the widely used coordinate transformation technique \cite{Pendry,Leonhardt,Tsang,ChenRotate,Rahm,Luo,SchurigRay}, a cylindrical wave expansion technique \cite{Ruan}, and a method based on the full-wave Mie scattering model \cite{Chen,LuoPRB} have also been applied. In addition, the full-wave finite element method (FEM) based commercial simulation software COMSOL Multiphysics$^{\rm TM}$ has been extensively used to model different cloaks and validate theoretical predictions \cite{Cummer,Cai,Cai2,ChenRotate,VasicPRB}, due to its ability of dealing with anisotropic and radial dependent material parameters. However similar to other frequency domain techniques, the FEM may become inefficient when wideband solutions are needed. So far, the time domain techniques that have been developed to model the cloaking structures include the time-dependent scattering theory \cite{Weder}, the transmission line method (TLM) \cite{BlanchardOE} and the finite-difference time domain (FDTD) method \cite{ZhaoOE}. Due to its simplicity in implementation and the ability of treating anisotropic, inhomogeneous and nonlinear materials, the FDTD method has been extremely popular for the analysis of electromagnetic structures. However due to the computational complexity, so far the FDTD modeling of three-dimensional (3-D) cloaking structures has not been attempted. In this paper, we extend our previously proposed 2-D FDTD method \cite{ZhaoOE} to the 3-D case and develop a parallel dispersive FDTD method to model 3-D cloaking structures and reveal the extraordinary behavior that is different from its counterparts under 2-D assumptions.

\section{Parallel dispersive FDTD modeling of 3-D cloaks}
A complete set of material parameters of the ideal cloak in spherical coordinate is given by \cite{Pendry}
\begin{eqnarray}
\varepsilon_r&\!\!\!=\!\!\!&\mu_r=\frac{R_2}{R_2-R_1}\left(\frac{r-R_1}{r}\right)^2,\nonumber\\
\varepsilon_\theta&\!\!\!=\!\!\!&\mu_\theta=\frac{R_2}{R_2-R_1},\nonumber\\
\varepsilon_\phi&\!\!\!=\!\!\!&\mu_\phi=\frac{R_2}{R_2-R_1},
\label{eq_parameter_ideal}
\end{eqnarray}
where $R_1$ and $R_2$ are the inner and outer radii of the cloak, respectively, and $r$ is the distance from a spatial point within the cloak to the center of the cloak. Since the values of $\varepsilon_r$ and $\mu_r$ are less than one (between $0$ and $\left(R_2-R_1\right)/R_2$), same as the case for left-handed materials (LHMs), the cloak cannot be directly modeled using the conventional FDTD method. However, one can map the material parameters using dispersive material models, for example, the Drude model
\begin{equation}
\varepsilon_r(\omega)=1-\frac{\omega^2_p}{\omega^2-j\omega\gamma},
\label{eq_Drude}
\end{equation}
where $\omega_p$ and $\gamma$ are the plasma and collision frequencies of the material, respectively. By varying the plasma frequency, the radial dependent material parameters in (\ref{eq_parameter_ideal}) can be achieved. For example, for the ideal lossless case considered in this paper, i.e. the collision frequency in (\ref{eq_Drude}) is equal to zero ($\gamma=0$), the radial dependent plasma frequency can be calculated using $\omega_p=\omega\sqrt{1-\varepsilon_r}$ with a given value of $\varepsilon_r$ calculated from (\ref{eq_parameter_ideal}). Note that in practice, the plasma frequency of the material depends on the periodicity of the split ring resonators (SRRs) \cite{Schurig} or wires \cite{Cai}, which varies along the radial direction. It should be noted that other dispersive material models (e.g. Debye, Lorentz etc.) can be also considered for the modeling of electromagnetic cloaks. However, the Drude model has the simplest form when implemented using the dispersive FDTD method and has been widely used in the modeling of metamaterials. The Lorentz model can be also used in FDTD simulations with some additional modifications to the iterative equations. The Debye model may be less accurate to characterize the dispersion behavior of metamaterials and rarely used in the community.

Since the conventional FDTD method \cite{Yee,Taflove} deals with frequency-independent materials, the frequency-dependent FDTD method is hence referred as the dispersive FDTD method \cite{Luebbers,Gandhi,Sullivan}. There are also different dispersive FDTD methods using different approaches to deal with the frequency-dependent material parameters: the recursive convolution (RC) method \cite{Luebbers}, the auxiliary differential equation (ADE) method \cite{Gandhi} and the $Z$-transform method \cite{Sullivan}. Due to its simplicity, we have chosen the ADE method to model 3-D cloaks in this paper.

The ADE dispersive FDTD method is based on the Faraday's and Ampere's Laws
\begin{eqnarray}
\nabla\times\textbf{E}&\!\!\!=&\!\!\!-\frac{\partial\textbf{B}}{\partial t},
\label{eq_Maxwell_E}\\
\nabla\times\textbf{H}&\!\!\!=&\!\!\!\frac{\partial\textbf{D}}{\partial t},
\label{eq_Maxwell_H}
\end{eqnarray}
as well as the constitutive relations $\textbf{D}=\varepsilon\textbf{E}$ and $\textbf{B}=\mu\textbf{H}$ where $\varepsilon$ and $\mu$ are expressed by (\ref{eq_parameter_ideal}). Equations (\ref{eq_Maxwell_E}) and (\ref{eq_Maxwell_H}) can be discretized following a standard procedure \cite{Yee,Taflove}, which leads to the conventional FDTD updating equations
\begin{eqnarray}
\textbf{B}^{n+1}&\!\!\!=&\!\!\!\textbf{B}^n-\Delta t\cdot\widetilde{\nabla}\times\textbf{E}^{n+\frac{1}{2}},
\label{eq_Maxwell_B_approx}\\
\textbf{D}^{n+1}&\!\!\!=&\!\!\!\textbf{D}^n+\Delta t\cdot\widetilde{\nabla}\times\textbf{H}^{n+\frac{1}{2}},
\label{eq_Maxwell_D_approx}
\end{eqnarray}
where $\widetilde{\nabla}$ is the discrete curl operator, $\Delta t$ is the discrete FDTD time step and $n$ is the number of the time steps.

In addition, auxiliary differential equations need to be taken into account and they can be discretized through the following steps. For the conventional Cartesian FDTD mesh, since the material parameters given in (\ref{eq_parameter_ideal}) are in spherical coordinates, the following coordinate transformation is used \cite{Bower}
\begin{eqnarray}
\left[\begin{array}{ccc}
\varepsilon_{xx} & \varepsilon_{xy} & \varepsilon_{xz}\\
\varepsilon_{yx} & \varepsilon_{yy} & \varepsilon_{yz}\\
\varepsilon_{zx} & \varepsilon_{zy} & \varepsilon_{zz}
\end{array}\right]&\!\!\!=&\!\!\!\left[\begin{array}{ccc}
\sin\theta\cos\phi & \cos\theta\cos\phi & -\sin\phi\\
\sin\theta\sin\phi & \cos\theta\sin\phi & \cos\phi\\
\cos\theta & -\sin\theta & 0
\end{array}\right]
\left[\begin{array}{ccc}
\varepsilon_{r} & 0 & 0\\
0 & \varepsilon_{\theta} & 0\\
0 & 0 & \varepsilon_{\phi}
\end{array}\right]\nonumber\\
&&\cdot\left[\begin{array}{ccc}
\sin\theta\cos\phi & \sin\theta\sin\phi & \cos\theta\\
\cos\theta\cos\phi & \cos\theta\sin\phi & -\sin\theta\\
-\sin\phi & \cos\phi & 0
\end{array}\right].
\label{eq_transform}
\end{eqnarray}
The tensor form of the constitutive relation is given by
\begin{eqnarray}
&&\varepsilon_0\left[\begin{array}{ccc}
\varepsilon_{xx} & \varepsilon_{xy} & \varepsilon_{xz}\\
\varepsilon_{yx} & \varepsilon_{yy} & \varepsilon_{yz}\\
\varepsilon_{zx} & \varepsilon_{zy} & \varepsilon_{zz}
\end{array}\right]\left[\begin{array}{c}
E_x\\E_y\\E_z\end{array}\right]=\left[\begin{array}{c}
D_x\\D_y\\D_z\end{array}\right]\nonumber\\
&\Leftrightarrow&\varepsilon_0\left[\begin{array}{c}
E_x\\E_y\\E_z\end{array}\right]=\left[\begin{array}{ccc}
\varepsilon_{xx} & \varepsilon_{xy} & \varepsilon_{xz}\\
\varepsilon_{yx} & \varepsilon_{yy} & \varepsilon_{yz}\\
\varepsilon_{zx} & \varepsilon_{zy} & \varepsilon_{zz}
\end{array}\right]^{-1}\left[\begin{array}{c}
D_x\\D_y\\D_z\end{array}\right],
\label{eq_constitutive}
\end{eqnarray}
where
\begin{equation}
\left[\begin{array}{ccc}
\varepsilon_{xx} & \varepsilon_{xy} & \varepsilon_{xz}\\
\varepsilon_{yx} & \varepsilon_{yy} & \varepsilon_{yz}\\
\varepsilon_{zx} & \varepsilon_{zy} & \varepsilon_{zz}
\end{array}\right]^{-1}=\left[\begin{array}{ccc}
\varepsilon'_{xx} & \varepsilon'_{xy} & \varepsilon'_{xz}\\
\varepsilon'_{yx} & \varepsilon'_{yy} & \varepsilon'_{yz}\\
\varepsilon'_{zx} & \varepsilon'_{zy} & \varepsilon'_{zz}
\end{array}\right],
\label{eq_inverse}
\end{equation}
and
\begin{eqnarray}
\varepsilon'_{xx}&\!\!=&\!\!\frac{1}{\varepsilon_r}\sin^2\theta\cos^2\phi+\frac{1}{\varepsilon_\theta}\cos^2\theta\cos^2\phi+\frac{1}{\varepsilon_\phi}\sin^2\phi,\nonumber\\
\varepsilon'_{xy}&\!\!=&\!\!\frac{1}{\varepsilon_r}\sin^2\theta\sin\phi\cos\phi+\frac{1}{\varepsilon_\theta}\cos^2\theta\sin\phi\cos\phi-\frac{1}{\varepsilon_\phi}\sin\phi\cos\phi,\nonumber\\
\varepsilon'_{xz}&\!\!=&\!\!\frac{1}{\varepsilon_r}\sin\theta\cos\theta\cos\phi-\frac{1}{\varepsilon_\theta}\sin\theta\cos\theta\cos\phi,\nonumber\\
\varepsilon'_{yx}&\!\!=&\!\!\frac{1}{\varepsilon_r}\sin^2\theta\sin\phi\cos\phi+\frac{1}{\varepsilon_\theta}\cos^2\theta\sin\phi\cos\phi-\frac{1}{\varepsilon_\phi}\sin\phi\cos\phi,\nonumber\\
\varepsilon'_{yy}&\!\!=&\!\!\frac{1}{\varepsilon_r}\sin^2\theta\sin^2\phi+\frac{1}{\varepsilon_\theta}\cos^2\theta\sin^2\phi+\frac{1}{\varepsilon_\phi}\cos^2\phi,\nonumber\\
\varepsilon'_{yz}&\!\!=&\!\!\frac{1}{\varepsilon_r}\sin\theta\cos\theta\sin\phi-\frac{1}{\varepsilon_\theta}\sin\theta\cos\theta\sin\phi,\nonumber\\
\varepsilon'_{zx}&\!\!=&\!\!\frac{1}{\varepsilon_r}\sin\theta\cos\theta\cos\phi-\frac{1}{\varepsilon_\theta}\sin\theta\cos\theta\cos\phi,\nonumber\\
\varepsilon'_{zy}&\!\!=&\!\!\frac{1}{\varepsilon_r}\sin\theta\cos\theta\sin\phi-\frac{1}{\varepsilon_\theta}\sin\theta\cos\theta\sin\phi,\nonumber\\
\varepsilon'_{zz}&\!\!=&\!\!\frac{1}{\varepsilon_r}\cos^2\theta+\frac{1}{\varepsilon_\theta}\sin^2\theta.
\label{eq_inverse_co}
\end{eqnarray}
Note that the inverse of the permittivity tensor matrix (\ref{eq_inverse}) exists only when $\varepsilon_r\neq0$, $\varepsilon_\theta\neq0$ and $\varepsilon_\phi\neq0$. However the inner boundary of the cloak does not satisfy the condition of $\varepsilon_r\neq0$. Therefore in our FDTD simulations, we place a perfect electric conductor (PEC) sphere with radius equal to $R_1$ inside the cloak to guarantee the validity of (\ref{eq_inverse}).

Substituting (\ref{eq_inverse}) into (\ref{eq_constitutive}) gives
\begin{eqnarray}
\varepsilon_0E_x&\!\!=&\!\!\left(\frac{1}{\varepsilon_r}\sin^2\theta\cos^2\phi+\frac{1}{\varepsilon_\theta}\cos^2\theta\cos^2\phi+\frac{1}{\varepsilon_\phi}\sin^2\phi\right)D_x\nonumber\\
&&+\left(\frac{1}{\varepsilon_r}\sin^2\theta\sin\phi\cos\phi+\frac{1}{\varepsilon_\theta}\cos^2\theta\sin\phi\cos\phi-\frac{1}{\varepsilon_\phi}\sin\phi\cos\phi\right)D_y\nonumber\\
&&+\left(\frac{1}{\varepsilon_r}\sin\theta\cos\theta\cos\phi-\frac{1}{\varepsilon_\theta}\sin\theta\cos\theta\cos\phi\right)D_z, \label{eq_constitutive1}\\
\varepsilon_0E_y&\!\!=&\!\!\left(\frac{1}{\varepsilon_r}\sin^2\theta\sin\phi\cos\phi+\frac{1}{\varepsilon_\theta}\cos^2\theta\sin\phi\cos\phi-\frac{1}{\varepsilon_\phi}\sin\phi\cos\phi\right)D_x\nonumber\\
&&+\left(\frac{1}{\varepsilon_r}\sin^2\theta\sin^2\phi+\frac{1}{\varepsilon_\theta}\cos^2\theta\sin^2\phi+\frac{1}{\varepsilon_\phi}\cos^2\phi\right)D_y\nonumber\\
&&+\left(\frac{1}{\varepsilon_r}\sin\theta\cos\theta\sin\phi-\frac{1}{\varepsilon_\theta}\sin\theta\cos\theta\sin\phi\right)D_z \label{eq_constitutive2}\\
\varepsilon_0E_z&\!\!=&\!\!\left(\frac{1}{\varepsilon_r}\sin\theta\cos\theta\cos\phi-\frac{1}{\varepsilon_\theta}\sin\theta\cos\theta\cos\phi\right)D_x\nonumber\\
&&+\left(\frac{1}{\varepsilon_r}\sin\theta\cos\theta\sin\phi-\frac{1}{\varepsilon_\theta}\sin\theta\cos\theta\sin\phi\right)D_y\nonumber\\
&&+\left(\frac{1}{\varepsilon_r}\cos^2\theta+\frac{1}{\varepsilon_\theta}\sin^2\theta\right)D_z.
\label{eq_constitutive3}
\end{eqnarray}
Since the above equations have a similar form, in the following, the derivation of the updating equation is only given for the $E_x$ component. The updating equations for the $E_y$ and $E_z$ components can be derived following the same procedure.

Express $\varepsilon_r$ in the Drude form of (\ref{eq_parameter_ideal}), Eq. (\ref{eq_constitutive1}) can be written as
\begin{eqnarray}
&&\varepsilon_0\left(\omega^2-j\omega\gamma-\omega^2_p\right)E_x=\Bigg[\left(\omega^2-j\omega\gamma\right)\sin^2\theta\cos^2\phi\nonumber\\
&&+\left.\left(\omega^2-j\omega\gamma-\omega^2_p\right)\left(\frac{\cos^2\theta\cos^2\phi}{\varepsilon_\theta}+\frac{\sin^2\phi}{\varepsilon_\phi}\right)\right]D_x\nonumber\\
&&+\left[\left(\omega^2-j\omega\gamma\right)\sin^2\theta\sin\phi\cos\phi+\left(\omega^2-j\omega\gamma-\omega^2_p\right)\frac{\cos^2\theta\sin\phi\cos\phi}{\varepsilon_\theta}\right.\nonumber\\
&&\left.-\left(\omega^2-j\omega\gamma-\omega^2_p\right)\frac{\sin\phi\cos\phi}{\varepsilon_\phi}\right]D_y+\Bigg[\left(\omega^2-j\omega\gamma\right)\sin\theta\cos\theta\cos\phi\nonumber\\
&&\left.-\left(\omega^2-j\omega\gamma-\omega^2_p\right)\frac{\sin\theta\cos\theta\cos\phi}{\varepsilon_\theta}\right]D_z.
\label{eq_Ex0}
\end{eqnarray}
Note that $\varepsilon_\theta$ and $\varepsilon_\phi$ remain in (\ref{eq_Ex0}) because their values are always greater than one and can be directly used in the conventional FDTD updating equations \cite{Yee,Taflove}. Applying the inverse Fourier transform and the following rules:
\begin{equation}
j\omega\rightarrow\frac{\partial}{\partial
t},\qquad\omega^2\rightarrow-\frac{\partial^2}{\partial t^2},
\label{eq_inverse_Fourier}
\end{equation}
Eq. (\ref{eq_Ex0}) can be rewritten in the time domain as
\begin{eqnarray}
&&\varepsilon_0\left(\frac{\partial^2}{\partial t^2}+\gamma\frac{\partial}{\partial t}+\omega^2_p\right)E_x=\left[\left(\frac{\partial^2}{\partial t^2}+\gamma\frac{\partial}{\partial t}\right)\sin^2\theta\cos^2\phi\right.\nonumber\\
&&\left.+\left(\frac{\partial^2}{\partial t^2}+\gamma\frac{\partial}{\partial t}+\omega^2_p\right)\frac{\cos^2\theta\cos^2\phi}{\varepsilon_\theta}+\left(\frac{\partial^2}{\partial t^2}+\gamma\frac{\partial}{\partial t}+\omega^2_p\right)\frac{\sin^2\phi}{\varepsilon_\phi}\right]D_x\nonumber\\
&&+\left[\left(\frac{\partial^2}{\partial t^2}+\gamma\frac{\partial}{\partial t}\right)\sin^2\theta\sin\phi\cos\phi+\left(\frac{\partial^2}{\partial t^2}+\gamma\frac{\partial}{\partial t}+\omega^2_p\right)\frac{\cos^2\theta\sin\phi\cos\phi}{\varepsilon_\theta}\right.\nonumber\\
&&\left.-\left(\frac{\partial^2}{\partial t^2}+\gamma\frac{\partial}{\partial t}+\omega^2_p\right)\frac{\sin\phi\cos\phi}{\varepsilon_\phi}\right]D_y+\left[\left(\frac{\partial^2}{\partial t^2}+\gamma\frac{\partial}{\partial t}\right)\sin\theta\cos\theta\cos\phi\right.\nonumber\\
&&\left.-\left(\frac{\partial^2}{\partial t^2}+\gamma\frac{\partial}{\partial t}+\omega^2_p\right)\frac{\sin\theta\cos\theta\cos\phi}{\varepsilon_\theta}\right]D_z.
\label{eq_Ex}
\end{eqnarray}

The FDTD simulation domain is represented by an equally spaced 3-D grid with the periods $\Delta x$, $\Delta y$ and $\Delta z$ along the $x$-, $y$- and $z$-directions, respectively. For the discretization of Eq. (\ref{eq_Ex}), we use the central finite difference operators in time ($\delta_t$ and $\delta^2_t$) and the central average operator with respect to time ($\mu_t$ and $\mu^2_t$):
\begin{eqnarray}
\frac{\partial^2}{\partial t^2}\rightarrow\frac{\delta^2_t}{(\Delta t)^2},\qquad\frac{\partial}{\partial t}\rightarrow\frac{\delta_t}{\Delta t}\mu_t,\qquad1\rightarrow\mu^2_t,\nonumber
\label{eq_operator}
\end{eqnarray}
where the operators $\delta_t$, $\delta^2_t$, $\mu_t$ and $\mu^2_t$ are defined as in \cite{Hildebrand}:
\begin{eqnarray}
\delta_t\textbf{F}|^n_{m_x,m_y,m_z}&\!\!\equiv&\!\!\textbf{F}|^{n+\frac{1}{2}}_{m_x,m_y,m_z}-\textbf{F}|^{n-\frac{1}{2}}_{m_x,m_y,m_z},\nonumber\\
\delta^2_t\textbf{F}|^n_{m_x,m_y,m_z}&\!\!\equiv&\!\!\textbf{F}|^{n+1}_{m_x,m_y,m_z}-2\textbf{F}|^n_{m_x,m_y,m_z}+\textbf{F}|^{n-1}_{m_x,m_y,m_z},\nonumber\\
\mu_t\textbf{F}|^n_{m_x,m_y,m_z}&\!\!\equiv&\!\!\frac{\textbf{F}|^{n+\frac{1}{2}}_{m_x,m_y,m_z}+\textbf{F}|^{n-\frac{1}{2}}_{m_x,m_y,m_z}}{2},\nonumber\\
\mu^2_t\textbf{F}|^n_{m_x,m_y,m_z}&\!\!\equiv&\!\!\frac{\textbf{F}|^{n+1}_{m_x,m_y,m_z}+2\textbf{F}|^n_{m_x,m_y,m_z}+\textbf{F}|^{n-1}_{m_x,m_y,m_z}}{4},
\label{eq_operators}
\end{eqnarray}
where $\textbf{F}$ represents field components and $m_x,m_y,m_z$ are the indices corresponding to a certain discretization point in the FDTD domain. The discretized Eq. (\ref{eq_Ex}) reads
\begin{eqnarray}
&&\varepsilon_0\left[\frac{\delta^2_t}{(\Delta t)^2}+\gamma\frac{\delta_t}{\Delta t}\mu_t+\omega^2_p\mu^2_t\right]E_x=\left\{\left[\frac{\delta^2_t}{(\Delta t)^2}+\gamma\frac{\delta_t}{\Delta t}\mu_t\right]\sin^2\theta\cos^2\phi\right.\nonumber\\
&&\left.+\left[\frac{\delta^2_t}{(\Delta t)^2}+\gamma\frac{\delta_t}{\Delta t}\mu_t+\omega^2_p\mu^2_t\right]\left(\frac{\cos^2\theta\cos^2\phi}{\varepsilon_\theta}+\frac{\sin^2\phi}{\varepsilon_\phi}\right)\right\}D_x\nonumber\\
&&+\left\{\left[\frac{\delta^2_t}{(\Delta t)^2}+\gamma\frac{\delta_t}{\Delta t}\mu_t\right]\sin^2\theta\sin\phi\cos\phi\right.\nonumber\\
&&\left.+\left[\frac{\delta^2_t}{(\Delta t)^2}+\gamma\frac{\delta_t}{\Delta t}\mu_t+\omega^2_p\mu^2_t\right]\left(\frac{\cos^2\theta\sin\phi\cos\phi}{\varepsilon_\theta}-\frac{\sin\phi\cos\phi}{\varepsilon_\phi}\right)\right\}D_y\nonumber\\
&&+\left\{\left[\frac{\delta^2_t}{(\Delta t)^2}+\gamma\frac{\delta_t}{\Delta t}\mu_t\right]\sin\theta\cos\theta\cos\phi\right.\nonumber\\
&&\left.-\left[\frac{\delta^2_t}{(\Delta t)^2}+\gamma\frac{\delta_t}{\Delta t}\mu_t+\omega^2_p\mu^2_t\right]\frac{\sin\theta\cos\theta\cos\phi}{\varepsilon_\theta}\right\}D_z.
\label{eq_Ex1}
\end{eqnarray}
Note that in (\ref{eq_Ex1}), the discretization of the term $\omega^2_p$ of (\ref{eq_Ex}) is performed using the central average operator $\mu^2_t$ in order to guarantee the improved stability; the central average operator $\mu_t$ is used for the term containing $\gamma$ to preserve the second-order feature of the equation. Equation (\ref{eq_Ex1}) can be written as
\begin{eqnarray}
&&\varepsilon_0\left[\frac{E^{n+1}_x-2E^n_x+E^{n-1}_x}{(\Delta t)^2}+\gamma\frac{E^{n+1}_x-E^{n-1}_x}{2\Delta
t}+\omega^2_p\frac{E^{n+1}_x+2E^n_x+E^{n-1}_x}{4}\right]\nonumber\\
&&=\sin^2\theta\cos^2\phi\left[\frac{D^{n+1}_x-2D^n_x+D^{n-1}_x}{(\Delta t)^2}+\gamma\frac{D^{n+1}_x-D^{n-1}_x}{2\Delta
t}\right]\nonumber\\
&&+\left(\frac{\cos^2\theta\cos^2\phi}{\varepsilon_\theta}+\frac{\sin^2\phi}{\varepsilon_\phi}\right)\left[\frac{D^{n+1}_x-2D^n_x+D^{n-1}_x}{(\Delta t)^2}+\gamma\frac{D^{n+1}_x-D^{n-1}_x}{2\Delta t}\right.\nonumber\\
&&\left.+\omega^2_p\frac{D^{n+1}_x+2D^n_x+D^{n-1}_x}{4}\right]+\sin^2\theta\sin\phi\cos\phi\left[\frac{D^{n+1}_y-2D^n_y+D^{n-1}_y}{(\Delta t)^2}\right.\nonumber\\
&&\left.+\gamma\frac{D^{n+1}_y-D^{n-1}_y}{2\Delta t}\right]+\left(\frac{\cos^2\theta\sin\phi\cos\phi}{\varepsilon_\theta}-\frac{\sin\phi\cos\phi}{\varepsilon_\phi}\right)\nonumber\\
&&\cdot\left[\frac{D^{n+1}_y-2D^n_y+D^{n-1}_y}{(\Delta t)^2}+\gamma\frac{D^{n+1}_y-D^{n-1}_y}{2\Delta t}+\omega^2_p\frac{D^{n+1}_y+2D^n_y+D^{n-1}_y}{4}\right]\nonumber\\
&&+\sin\theta\cos\theta\cos\phi\left[\frac{D^{n+1}_z-2D^n_z+D^{n-1}_z}{(\Delta t)^2}+\gamma\frac{D^{n+1}_z-D^{n-1}_z}{2\Delta
t}\right]\nonumber\\
&&-\frac{\sin\theta\cos\theta\cos\phi}{\varepsilon_\theta}\left[\frac{D^{n+1}_z-2D^n_z+D^{n-1}_z}{(\Delta t)^2}+\gamma\frac{D^{n+1}_z-D^{n-1}_z}{2\Delta t}\right.\nonumber\\
&&\left.+\omega^2_p\frac{D^{n+1}_z+2D^n_z+D^{n-1}_z}{4}\right].
\label{eq_Ex2}
\end{eqnarray}
After simple manipulations, the updating equation for $E_x$ can be obtained as
\begin{eqnarray}
E^{n+1}_x&\!\!\!=&\!\!\!\! \left[b_{0xx}D^{n+1}_x+b_{1xx}D^n_x+b_{2xx}D^{n-1}_x+b_{0xy}\overline{D_y}^{n+1}+b_{1xy}\overline{D_y}^n+b_{2xy}\overline{D_y}^{n-1}\right.\nonumber\\
&&\!\!\!+b_{0xz}\overline{D_z}^{n+1}+b_{1xz}\overline{D_z}^n+b_{2xz}\overline{D_z}^{n-1}-\left(a_{1x}E^n_x+a_{2x}E^{n-1}_x\right)\bigg]/a_{0x},
\label{eq_Ex3}
\end{eqnarray}
where the coefficients are given by
{\footnotesize\begin{eqnarray}
a_{0x}&\!\!\!=&\!\!\!\varepsilon_0\left[\frac{1}{(\Delta t)^2}+\frac{\gamma}{2\Delta t}+\frac{\omega^2_p}{4}\right],~~~a_{1x}=\varepsilon_0\left[-\frac{2}{(\Delta t)^2}+\frac{\omega^2_p}{2}\right],\nonumber\\
a_{2x}&\!\!\!=&\!\!\!\varepsilon_0\left[\frac{1}{(\Delta t)^2}-\frac{\gamma}{2\Delta t}+\frac{\omega^2_p}{4}\right],\nonumber\\
b_{0xx}&\!\!\!=&\!\!\!\sin^2\theta\cos^2\phi\left[\frac{1}{(\Delta t)^2}+\frac{\gamma}{2\Delta t}\right]+\left(\frac{\cos^2\theta\cos^2\phi}{\varepsilon_\theta}+\frac{\sin^2\phi}{\varepsilon_\phi}\right)\left[\frac{1}{(\Delta t)^2}+\frac{\gamma}{2\Delta t}+\frac{\omega^2_p}{4}\right],\nonumber\\
b_{1xx}&\!\!\!=&\!\!\!-\sin^2\theta\cos^2\phi\frac{2}{(\Delta t)^2}+\left(\frac{\cos^2\theta\cos^2\phi}{\varepsilon_\theta}+\frac{\sin^2\phi}{\varepsilon_\phi}\right)\left[-\frac{2}{(\Delta t)^2}+\frac{\omega^2_p}{2}\right],\nonumber\\
b_{2xx}&\!\!\!=&\!\!\!\sin^2\theta\cos^2\phi\left[\frac{1}{(\Delta t)^2}-\frac{\gamma}{2\Delta t}\right]+\left(\frac{\cos^2\theta\cos^2\phi}{\varepsilon_\theta}+\frac{\sin^2\phi}{\varepsilon_\phi}\right)\left[\frac{1}{(\Delta t)^2}-\frac{\gamma}{2\Delta t}+\frac{\omega^2_p}{4}\right],\nonumber\\
b_{0xy}&\!\!\!=&\!\!\!\sin^2\theta\sin\phi\cos\phi\left[\frac{1}{(\Delta t)^2}+\frac{\gamma}{2\Delta t}\right]+\left(\frac{\cos^2\theta\sin\phi\cos\phi}{\varepsilon_\theta}-\frac{\sin\phi\cos\phi}{\varepsilon_\phi}\right)\nonumber\\
&&\!\!\!\cdot\left[\frac{1}{(\Delta t)^2}+\frac{\gamma}{2\Delta t}+\frac{\omega^2_p}{4}\right],\nonumber\\
b_{1xy}&\!\!\!=&\!\!\!-\sin^2\theta\sin\phi\cos\phi\frac{2}{(\Delta t)^2}+\left(\frac{\cos^2\theta\sin\phi\cos\phi}{\varepsilon_\theta}-\frac{\sin\phi\cos\phi}{\varepsilon_\phi}\right)\left[-\frac{2}{(\Delta t)^2}+\frac{\omega^2_p}{2}\right],\nonumber\\
b_{2xy}&\!\!\!=&\!\!\!\sin^2\theta\sin\phi\cos\phi\left[\frac{1}{(\Delta t)^2}-\frac{\gamma}{2\Delta t}\right]+\left(\frac{\cos^2\theta\sin\phi\cos\phi}{\varepsilon_\theta}-\frac{\sin\phi\cos\phi}{\varepsilon_\phi}\right)\nonumber\\
&&\!\!\!\cdot\left[\frac{1}{(\Delta t)^2}-\frac{\gamma}{2\Delta t}+\frac{\omega^2_p}{4}\right],\nonumber\\
b_{0xz}&\!\!\!=&\!\!\!\sin\theta\cos\theta\cos\phi\left[\frac{1}{(\Delta t)^2}+\frac{\gamma}{2\Delta t}\right]-\frac{\sin\theta\cos\theta\cos\phi}{\varepsilon_\theta}\left[\frac{1}{(\Delta t)^2}+\frac{\gamma}{2\Delta t}+\frac{\omega^2_p}{4}\right],\nonumber\\
b_{1xz}&\!\!\!=&\!\!\!-\sin\theta\cos\theta\cos\phi\frac{2}{(\Delta t)^2}-\frac{\sin\theta\cos\theta\cos\phi}{\varepsilon_\theta}\left[-\frac{2}{(\Delta t)^2}+\frac{\omega^2_p}{2}\right],\nonumber\\
b_{2xz}&\!\!\!=&\!\!\!\sin\theta\cos\theta\cos\phi\left[\frac{1}{(\Delta t)^2}-\frac{\gamma}{2\Delta t}\right]-\frac{\sin\theta\cos\theta\cos\phi}{\varepsilon_\theta}\left[\frac{1}{(\Delta t)^2}-\frac{\gamma}{2\Delta t}+\frac{\omega^2_p}{4}\right].\nonumber
\end{eqnarray}}
Following the same procedure, the updating equation for $E_y$ is
\begin{eqnarray}
E^{n+1}_y&\!\!\!=&\!\!\!\! \left[b_{0yx}\overline{D_x}^{n+1}+b_{1yx}\overline{D_x}^n+b_{2yx}\overline{D_x}^{n-1}+b_{0yy}D^{n+1}_y+b_{1yy}D^n_y+b_{2yy}D^{n-1}_y\right.\nonumber\\
&&\!\!\!+b_{0yz}\overline{D_z}^{n+1}+b_{1yz}\overline{D_z}^n+b_{2yz}\overline{D_z}^{n-1}-\left(a_{1y}E^n_y+a_{2y}E^{n-1}_y\right)\bigg]/a_{0y},
\label{eq_Ey}
\end{eqnarray}
with the coefficients given by
{\footnotesize\begin{eqnarray}
a_{0y}&\!\!\!=&\!\!\!\varepsilon_0\left[\frac{1}{(\Delta t)^2}+\frac{\gamma}{2\Delta t}+\frac{\omega^2_p}{4}\right],~~~a_{1y}=\varepsilon_0\left[-\frac{2}{(\Delta t)^2}+\frac{\omega^2_p}{2}\right],\nonumber\\
a_{2y}&\!\!\!=&\!\!\!\varepsilon_0\left[\frac{1}{(\Delta t)^2}-\frac{\gamma}{2\Delta t}+\frac{\omega^2_p}{4}\right],\nonumber\\
b_{0yx}&\!\!\!=&\!\!\!\sin^2\theta\sin\phi\cos\phi\left[\frac{1}{(\Delta t)^2}+\frac{\gamma}{2\Delta t}\right]+\left(\frac{\cos^2\theta\sin\phi\cos\phi}{\varepsilon_\theta}-\frac{\sin\phi\cos\phi}{\varepsilon_\phi}\right)\nonumber\\
&&\!\!\!\cdot\left[\frac{1}{(\Delta t)^2}+\frac{\gamma}{2\Delta t}+\frac{\omega^2_p}{4}\right],\nonumber\\
b_{1yx}&\!\!\!=&\!\!\!-\sin^2\theta\sin\phi\cos\phi\frac{2}{(\Delta t)^2}+\left(\frac{\cos^2\theta\sin\phi\cos\phi}{\varepsilon_\theta}-\frac{\sin\phi\cos\phi}{\varepsilon_\phi}\right)\left[-\frac{2}{(\Delta t)^2}+\frac{\omega^2_p}{2}\right],\nonumber\\
b_{2yx}&\!\!\!=&\!\!\!\sin^2\theta\sin\phi\cos\phi\left[\frac{1}{(\Delta t)^2}-\frac{\gamma}{2\Delta t}\right]+\left(\frac{\cos^2\theta\sin\phi\cos\phi}{\varepsilon_\theta}-\frac{\sin\phi\cos\phi}{\varepsilon_\phi}\right)\nonumber\\
&&\!\!\!\cdot\left[\frac{1}{(\Delta t)^2}-\frac{\gamma}{2\Delta t}+\frac{\omega^2_p}{4}\right],\nonumber\\
b_{0yy}&\!\!\!=&\!\!\!\sin^2\theta\sin^2\phi\left[\frac{1}{(\Delta t)^2}+\frac{\gamma}{2\Delta t}\right]+\left(\frac{\cos^2\theta\sin^2\phi}{\varepsilon_\theta}+\frac{\cos^2\phi}{\varepsilon_\phi}\right)\left[\frac{1}{(\Delta t)^2}+\frac{\gamma}{2\Delta t}+\frac{\omega^2_p}{4}\right],\nonumber\\
b_{1yy}&\!\!\!=&\!\!\!-\sin^2\theta\sin^2\phi\frac{2}{(\Delta t)^2}+\left(\frac{\cos^2\theta\sin^2\phi}{\varepsilon_\theta}+\frac{\cos^2\phi}{\varepsilon_\phi}\right)\left[-\frac{2}{(\Delta t)^2}+\frac{\omega^2_p}{2}\right],\nonumber\\
b_{2yy}&\!\!\!=&\!\!\!\sin^2\theta\sin^2\phi\left[\frac{1}{(\Delta t)^2}-\frac{\gamma}{2\Delta t}\right]+\left(\frac{\cos^2\theta\sin^2\phi}{\varepsilon_\theta}+\frac{\cos^2\phi}{\varepsilon_\phi}\right)\left[\frac{1}{(\Delta t)^2}-\frac{\gamma}{2\Delta t}+\frac{\omega^2_p}{4}\right],\nonumber\\
b_{0yz}&\!\!\!=&\!\!\!\sin\theta\cos\theta\sin\phi\left[\frac{1}{(\Delta t)^2}+\frac{\gamma}{2\Delta t}\right]-\frac{\sin\theta\cos\theta\sin\phi}{\varepsilon_\theta}\left[\frac{1}{(\Delta t)^2}+\frac{\gamma}{2\Delta t}+\frac{\omega^2_p}{4}\right],\nonumber\\
b_{1yz}&\!\!\!=&\!\!\!-\sin\theta\cos\theta\sin\phi\frac{2}{(\Delta t)^2}-\frac{\sin\theta\cos\theta\sin\phi}{\varepsilon_\theta}\left[-\frac{2}{(\Delta t)^2}+\frac{\omega^2_p}{2}\right],\nonumber\\
b_{2yz}&\!\!\!=&\!\!\!\sin\theta\cos\theta\sin\phi\left[\frac{1}{(\Delta t)^2}-\frac{\gamma}{2\Delta t}\right]-\frac{\sin\theta\cos\theta\sin\phi}{\varepsilon_\theta}\left[\frac{1}{(\Delta t)^2}-\frac{\gamma}{2\Delta t}+\frac{\omega^2_p}{4}\right].\nonumber
\end{eqnarray}}
And the updating equations for $E_z$ is
\begin{eqnarray}
E^{n+1}_z&\!\!\!=&\!\!\! \left[b_{0zx}\overline{D_x}^{n+1}+b_{1zx}\overline{D_x}^n+b_{2zx}\overline{D_x}^{n-1}+b_{0zy}\overline{D_y}^{n+1}+b_{1zy}\overline{D_y}^n\right.\nonumber\\
&&\!\!\!+b_{2zy}\overline{D_y}^{n-1}+b_{0zz}D^{n+1}_z+b_{1zz}D^n_z+b_{2zz}D^{n-1}_z\nonumber\\
&&\!\!\!-\left(a_{1z}E^n_z+a_{2z}E^{n-1}_z\right)\bigg]/a_{0z},
\label{eq_Ez}
\end{eqnarray}
with the coefficients given by
{\footnotesize\begin{eqnarray}
a_{0z}&\!\!\!=&\!\!\!\varepsilon_0\left[\frac{1}{(\Delta t)^2}+\frac{\gamma}{2\Delta t}+\frac{\omega^2_p}{4}\right],~~~a_{1z}=\varepsilon_0\left[-\frac{2}{(\Delta t)^2}+\frac{\omega^2_p}{2}\right],\nonumber\\
a_{2z}&\!\!\!=&\!\!\!\varepsilon_0\left[\frac{1}{(\Delta t)^2}-\frac{\gamma}{2\Delta t}+\frac{\omega^2_p}{4}\right],\nonumber\\
b_{0zx}&\!\!\!=&\!\!\!\sin\theta\cos\theta\cos\phi\left[\frac{1}{(\Delta t)^2}+\frac{\gamma}{2\Delta t}\right]-\frac{\sin\theta\cos\theta\cos\phi}{\varepsilon_\theta}\left[\frac{1}{(\Delta t)^2}+\frac{\gamma}{2\Delta t}+\frac{\omega^2_p}{4}\right],\nonumber\\
b_{1zx}&\!\!\!=&\!\!\!-\sin\theta\cos\theta\cos\phi\frac{2}{(\Delta t)^2}-\frac{\sin\theta\cos\theta\cos\phi}{\varepsilon_\theta}\left[-\frac{2}{(\Delta t)^2}+\frac{\omega^2_p}{2}\right],\nonumber\\
b_{2zx}&\!\!\!=&\!\!\!\sin\theta\cos\theta\cos\phi\left[\frac{1}{(\Delta t)^2}-\frac{\gamma}{2\Delta t}\right]-\frac{\sin\theta\cos\theta\cos\phi}{\varepsilon_\theta}\left[\frac{1}{(\Delta t)^2}-\frac{\gamma}{2\Delta t}+\frac{\omega^2_p}{4}\right],\nonumber\\
b_{0zy}&\!\!\!=&\!\!\!\sin\theta\cos\theta\sin\phi\left[\frac{1}{(\Delta t)^2}+\frac{\gamma}{2\Delta t}\right]-\frac{\sin\theta\cos\theta\sin\phi}{\varepsilon_\theta}\left[\frac{1}{(\Delta t)^2}+\frac{\gamma}{2\Delta t}+\frac{\omega^2_p}{4}\right],\nonumber\\
b_{1zy}&\!\!\!=&\!\!\!-\sin\theta\cos\theta\sin\phi\frac{2}{(\Delta t)^2}-\frac{\sin\theta\cos\theta\sin\phi}{\varepsilon_\theta}\left[-\frac{2}{(\Delta t)^2}+\frac{\omega^2_p}{2}\right],\nonumber\\
b_{2zy}&\!\!\!=&\!\!\!\sin\theta\cos\theta\sin\phi\left[\frac{1}{(\Delta t)^2}-\frac{\gamma}{2\Delta t}\right]-\frac{\sin\theta\cos\theta\sin\phi}{\varepsilon_\theta}\left[\frac{1}{(\Delta t)^2}-\frac{\gamma}{2\Delta t}+\frac{\omega^2_p}{4}\right],\nonumber\\
b_{0zz}&\!\!\!=&\!\!\!\cos^2\theta\left[\frac{1}{(\Delta t)^2}+\frac{\gamma}{2\Delta t}\right]+\frac{\sin^2\theta}{\varepsilon_\theta}\left[\frac{1}{(\Delta t)^2}+\frac{\gamma}{2\Delta t}+\frac{\omega^2_p}{4}\right],\nonumber\\
b_{1zz}&\!\!\!=&\!\!\!-\cos^2\theta\frac{2}{(\Delta t)^2}+\frac{\sin^2\theta}{\varepsilon_\theta}\left[-\frac{2}{(\Delta t)^2}+\frac{\omega^2_p}{2}\right],\nonumber\\
b_{2zz}&\!\!\!=&\!\!\!\cos^2\theta\left[\frac{1}{(\Delta t)^2}-\frac{\gamma}{2\Delta t}\right]+\frac{\sin^2\theta}{\varepsilon_\theta}\left[\frac{1}{(\Delta t)^2}-\frac{\gamma}{2\Delta t}+\frac{\omega^2_p}{4}\right].\nonumber
\end{eqnarray}}

Note that the field quantities $\overline{D_x}$, $\overline{D_y}$ and $\overline{D_z}$ in (\ref{eq_Ex3})-(\ref{eq_Ez}) are locally averaged values of $D_x$, $D_y$ and $D_z$, respectively since the $x$-, $y$- and $z$-components of the electric fields are in different locations in the FDTD domain \cite{Lee}. However, the averaging needs to be applied along different directions depending on the updating equations. Specifically in (\ref{eq_Ex3}), the averaged $D_y$ and $D_z$ can be calculated using
\begin{eqnarray}
\overline{D_y}(i,j,k)&\!\!\!=&\!\!\!\frac{D_y(i,j,k)\!+\!D_y(i+1,j,k)\!+\!D_y(i,j-1,k)\!+\!D_y(i+1,j-1,k)}{4},\nonumber\\
\overline{D_z}(i,j,k)&\!\!\!=&\!\!\!\frac{D_z(i,j,k)\!+\!D_z(i+1,j,k)\!+\!D_z(i,j,k-1)\!+\!D_z(i+1,j,k-1)}{4},\nonumber
\end{eqnarray}
where $(i,j,k)$ is the coordinate of the field component. In (\ref{eq_Ey}), the averaged $D_x$ and $D_z$ can be calculated using
\begin{eqnarray}
\overline{D_x}(i,j,k)&\!\!\!=&\!\!\!\frac{D_x(i,j,k)\!+\!D_x(i,j+1,k)\!+\!D_x(i-1,j,k)\!+\!D_x(i-1,j+1,k)}{4},\nonumber\\
\overline{D_z}(i,j,k)&\!\!\!=&\!\!\!\frac{D_z(i,j,k)\!+\!D_z(i,j+1,k)\!+\!D_z(i,j,k-1)\!+\!D_z(i,j+1,k-1)}{4}.\nonumber
\end{eqnarray}
And in (\ref{eq_Ez}), the averaged $D_x$ and $D_y$ can be calculated using
\begin{eqnarray}
\overline{D_x}(i,j,k)&\!\!\!=&\!\!\!\frac{D_x(i,j,k)\!+\!D_x(i,j,k+1)\!+\!D_x(i-1,j,k)\!+\!D_x(i-1,j,k+1)}{4},\nonumber\\
\overline{D_y}(i,j,k)&\!\!\!=&\!\!\!\frac{D_y(i,j,k)\!+\!D_y(i,j,k+1)\!+\!D_y(i,j-1,k)\!+\!D_y(i,j-1,k+1)}{4}.\nonumber
\end{eqnarray}

The updating equations for the magnetic fields $H_x$, $H_y$ and $H_z$ are in the same form as (\ref{eq_Ex3})-(\ref{eq_Ez}) with the same coefficients, and can be obtained by replacing $\textbf{E}$ with $\textbf{H}$ and $\textbf{D}$ with $\textbf{B}$. The averaged field components can be calculated in a similar manner. Equations (\ref{eq_Maxwell_B_approx}), (\ref{eq_Maxwell_D_approx}), (\ref{eq_Ex3})-(\ref{eq_Ez}) and the updating equations for $\textbf{H}$ from $\textbf{B}$ (not given) form the updating equation set for the modeling of 3-D cloaks using the well-known leap-frog scheme \cite{Yee}. If the plasma frequency in (\ref{eq_Drude}) is equal to zero i.e. $\omega_p=0$, and $\varepsilon_\theta=\mu_\theta=\varepsilon_\phi=\mu_\phi=1$, the above updating equation set reduces to the updating equation set for the free space.

Since the FDTD method is inherently a numerical technique, the spatial as well as time discretization has important effects on the accuracy of simulation results. Also because the permittivity and permeability are frequency dependent, one can expect a slight difference between the analytical and numerical material parameters due to the discrete time step in the FDTD method. From our previous analysis \cite{ZhaoJOPA} that for the modeling of metamaterials especially the LHMs, the numerical errors due to the time discretization will cause spurious resonances, hence a requirement of $\Delta x<\lambda/80$ is necessary. Following the same approach as in \cite{ZhaoOE}, one can find that the numerical permittivity $\widetilde{\varepsilon_r}$ for the ideal 3-D cloak takes the following form:
\begin{equation}
\widetilde{\varepsilon_r}=\varepsilon_0\left[1-\frac{\omega^2_p(\Delta t)^2\cos^2\frac{\omega\Delta t}{2}}{2\sin\frac{\omega\Delta t}{2}\left(2\sin\frac{\omega\Delta t}{2}-j\gamma\Delta t\cos\frac{\omega\Delta t}{2}\right)}\right].
\label{eq_epsn}
\end{equation}
Note that Eq. (\ref{eq_epsn}) simplifies to the analytical Drude dispersion model (\ref{eq_Drude}) when $\Delta t\rightarrow0$. With the expression of the numerical permittivity (\ref{eq_epsn}) available, one can correct the errors introduced by the discrepancy between the numerical and analytical material parameters. For example, if the required permittivity is $\varepsilon_r=\varepsilon'_r+j\varepsilon''_r$, after simple derivations, the corrected plasma and collision frequencies can be calculated as
\begin{eqnarray}
\widetilde{\omega_p}^2&\!\!\!=&\!\!\!\frac{2\sin\frac{\omega\Delta t}{2}\left[-2(\varepsilon'_r-1)\sin\frac{\omega\Delta t}{2}-\varepsilon''_r\gamma\Delta t\cos\frac{\omega\Delta t}{2}\right]}{(\Delta t)^2\cos^2\frac{\omega\Delta t}{2}},\nonumber\\
\widetilde{\gamma}&\!\!\!=&\!\!\!\frac{2\varepsilon''_r\sin\frac{\omega\Delta t}{2}}{(\varepsilon'_r-1)\Delta t\cos\frac{\omega\Delta t}{2}}.
\label{eq_corrected}
\end{eqnarray}

The above averaging of field components and the correction of numerical material parameters ensure stable and accurate FDTD simulations of the 3-D cloak. However if the averaging is not applied, the field distribution becomes unsymmetrical and hence incorrect; if the numerical material parameters are not corrected, the FDTD simulations become unstable before reaching the steady-state. Therefore in our simulations of the 3-D cloaks, the field averaging and corrected material parameters (\ref{eq_corrected}) are always used.

The FDTD method is a versatile numerical technique. However, similar to other numerical methods, it is computationally intensive. For large electromagnetic problems such as the modeling of 3-D cloaks, the requirement for system resources is beyond the capability of a single personal computer (PC). One way to resolve this problem is to divide the whole computational domain into many smaller sub-domains, and each sub-domain can be handled by a PC. By linking the PCs together with an appropriate synchronization procedure, the original large problem can be decomposed and solved efficiently.

\begin{figure}[htbp]
\centering
\includegraphics[width=10cm]{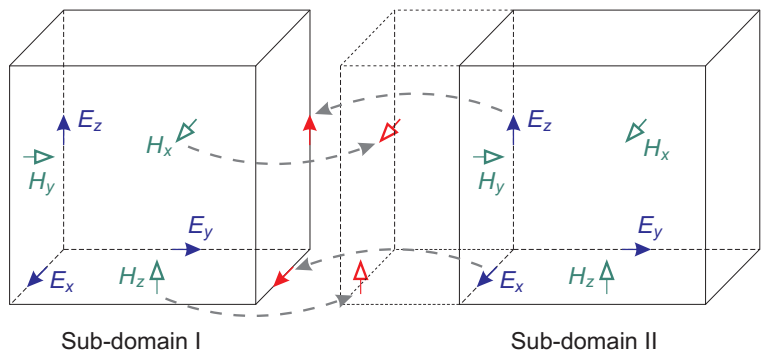}
\caption{The arrangement of field components in different sub-domains in parallel FDTD simulations. The red arrows indicate the field components which are transferred from the neighboring domain during the data communication process and used to update the field components on the boundary of the current sub-domain.}
\label{fig_pfdtd}
\end{figure}

One of the most attractive features of the FDTD method is that it can be easily parallelized with very little modifications to the algorithm. Since it solves Maxwell equations in the time-space domain, the parallel FDTD algorithm is based on the space decomposition technique \cite{Chew,Gedney}. The data transfer functionality between processors (PCs) is provided by the message passing interface (MPI) library. Data exchange is required only for the adjacent cells at the interface between different sub-domains and is performed at each time step, hence the parallel FDTD algorithm is a self-synchronized process. Figure~\ref{fig_pfdtd} shows the arrangement of the field components in different sub-domains in parallel FDTD simulations.
The red arrows are the transferred field components from the neighboring sub-domain during the data communication process. At the end of parallel FDTD simulations, the results calculated at each processor need to be combined to obtain the final simulation result. In comparison to conventional parallel FDTD method, the parallelization of the dispersive FDTD method introduced in this paper requires additional field components to be transferred between adjacent sub-domains during the synchronization process, because of the applied field averaging scheme. The complexity of algorithm further increases if the whole computational domain is divided along more than one direction, although the data communication load and the overall simulation time may be reduced.

The PC cluster used to simulate 3-D cloaks in Queen Mary, University of London consists of one head node for monitoring purposes and fifteen compute nodes for performing calculation tasks. Each node has Dual Intel Xeon E5405 (Quad Core 2.0 GHz) central processing units (CPUs) and there are 128 cores and 512 GB memory in total. The nodes are connected by a 24-port gigabit switch. The GNU C compiler (GCC) and a free version of MPI, MPI Chameleon (MPICH), developed by Argonne National Laboratory \cite{MPICH}, are used to compile the developed parallel dispersive FDTD code and handle the inter-core data communications, respectively. The above developed parallel dispersive FDTD method has been implemented to model the ideal 3-D cloaks, and the simulation results and discussions are presented in the following section.

\section{Numerical results and discussions}
\begin{figure}[htbp]
\centering
\includegraphics[width=10.5cm]{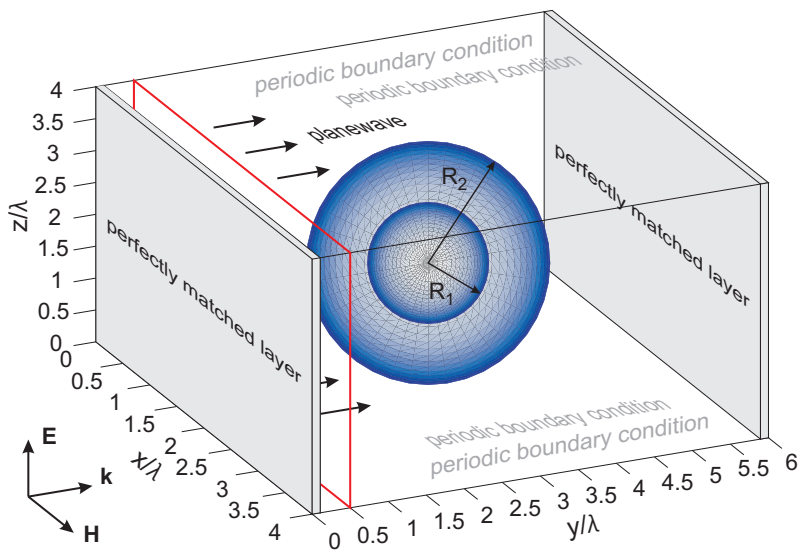}
\caption{The 3-D parallel dispersive FDTD simulation domain for the case of plane-wave incidence on the cloak. The red rectangle indicates the location of the source plane.}
\label{fig_domain}
\end{figure}

The 3-D FDTD simulation domain is shown in Fig.~\ref{fig_domain}.
The FDTD cell size in all simulations is $\Delta x=\Delta y=\lambda/150$ where $\lambda$ is the wavelength at the operating frequency $f=2.0$ GHz. The time step is chosen according to the Courant stability criterion \cite{Taflove} i.e. $\Delta t=\Delta x/\sqrt{3}c$. The radii of the cloak are $R_1=0.1$ m and $R_2=0.2$ m. In the present paper, only the ideal case (lossless) is considered i.e. the collision frequency in (\ref{eq_Drude}) is equal to zero ($\gamma=0$). The radial dependent plasma frequency can be calculated using (\ref{eq_corrected}) with a given value of $\varepsilon_r$ calculated from (\ref{eq_parameter_ideal}). The computational domain is truncated using Berenger's perfectly matched layer (PML) \cite{Berenger} in $y$-direction to absorb waves leaving the computational domain without introducing reflections, and terminated with periodic boundary conditions (PBCs) in $x$- and $z$-directions for the modeling of a plane-wave source. The plane-wave source is implemented by specifying a complete plane of FDTD cells using a certain wave function. The electric and magnetic fields of the plane wave are along the $z$- and $x$-axis, respectively and the propagation direction is along the $y$-axis, as indicated in Fig.~\ref{fig_domain}. For simplicity, the whole simulation domain is only divided along $y$-direction into 100 sub-domains and in total 100 processors and 220 gigabyte (GB) memory were used to run the parallel dispersive FDTD simulations. Each simulation lasts around 45 hours (13000 time steps) before reaching the steady-state.

\begin{figure}[htbp]
\centering
\includegraphics[width=12cm]{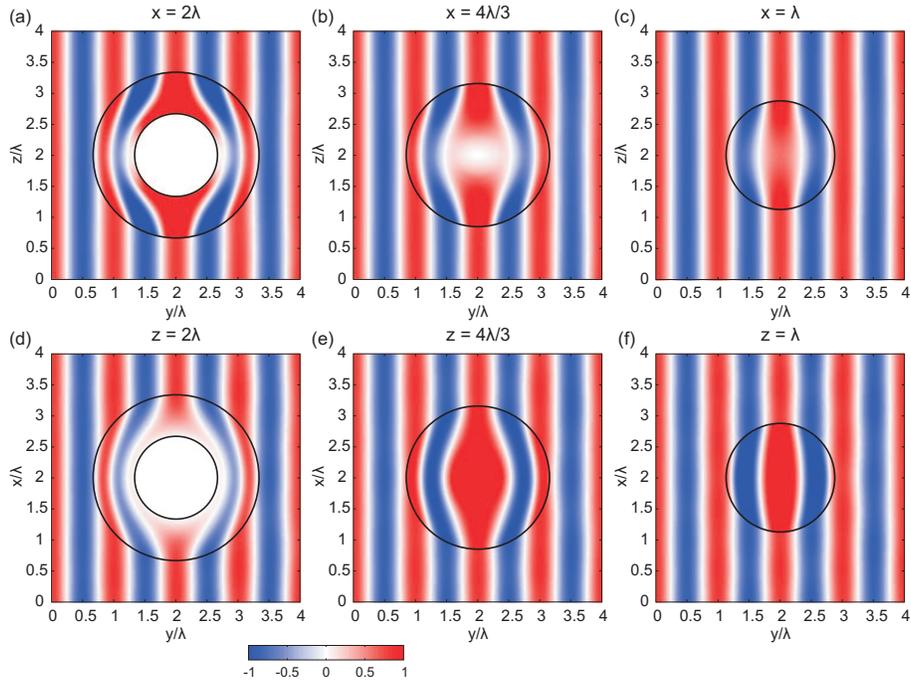}
\caption{Normalized field distributions for the $E_z$ component in (a)-(c) $y$-$z$ plane and (d)-(f) $x$-$y$ plane in the steady-state of the parallel dispersive FDTD simulations. The cutting planes are (see Fig.~\ref{fig_domain}): (a) $x=2\lambda$, (b) $x=4\lambda/3$, (c) $x=\lambda$, (d) $z=2\lambda$, (e) $z=4\lambda/3$, (f) $z=\lambda$. The wave propagation direction is from left to right.}
\label{fig_Ex}
\end{figure}

\begin{figure}[htbp]
\centering
\includegraphics[width=12cm]{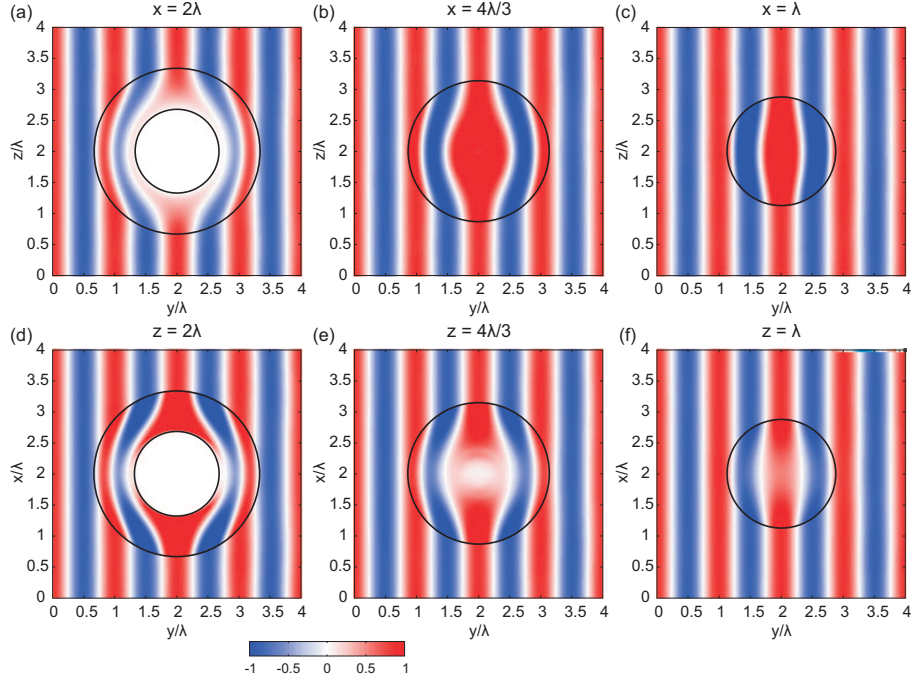}
\caption{Normalized field distributions for the $H_x$ component in (a)-(c) $y$-$z$ plane and (d)-(f) $x$-$y$ plane in the steady-state of the parallel dispersive FDTD simulations. The cutting planes are (see Fig.~\ref{fig_domain}): (a) $x=2\lambda$, (b) $x=4\lambda/3$, (c) $x=\lambda$, (d) $z=2\lambda$, (e) $z=4\lambda/3$, (f) $z=\lambda$. The wave propagation direction is from left to right.}
\label{fig_Hy}
\end{figure}

\begin{figure}[htbp]
\centering
\includegraphics[width=7cm]{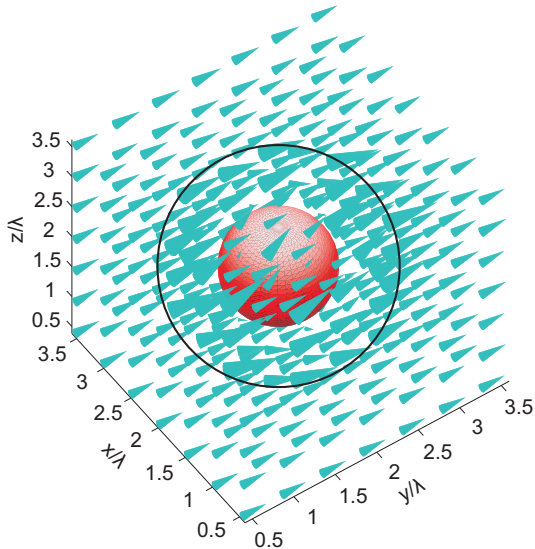}
\caption{Power flow diagram of a plane wave incidence on the ideal 3-D cloak calculated from parallel dispersive FDTD simulations.}
\label{fig_powerflow}
\end{figure}

Figures~\ref{fig_Ex} and \ref{fig_Hy} show the normalized steady-state field distributions for the $E_z$ and $H_x$ components in $y$-$z$ and $x$-$y$ planes, respectively.
It can be seen that the plane wave is guided by the cloak to propagate around its central region, and recomposed back after leaving the cloak. There is nearly no reflection (except those tiny numerical ones due to the finite spatial resolution in FDTD simulations), since the material parameters (\ref{eq_parameter_ideal}) vary continuously in space while keeping the impedance the same as the free space one. It is also interesting to notice that the $E_z$ component in $y$-$z$ and $x$-$y$ planes in Fig.~\ref{fig_Ex} and the $H_x$ component in $x$-$y$ and $y$-$z$ planes in Fig.~\ref{fig_Hy} have the same distributions (with different amplitude), which is due to the fact that the ideal 3-D cloak is a rotationally symmetric structure with respect to the electric and magnetic fields. The wave behavior near the 3-D cloak can be better illustrated using the power flow diagram, as plotted in Fig.~\ref{fig_powerflow}.
It is shown that the Poynting vectors are diverted around the central area enclosed by the cloak. Therefore objects placed inside the cloak do not introduce any scattering to external radiations and hence become `invisible'.

\begin{figure}[htbp]
\centering
\includegraphics[width=12cm]{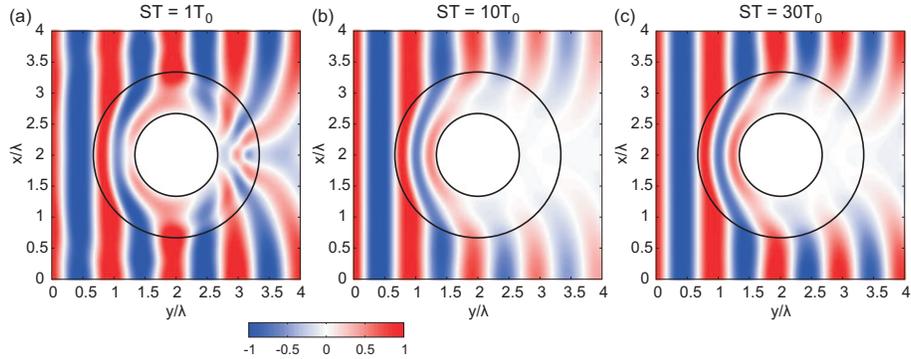}
\caption{Comparison of the influence of different switching time (ST) of the sinusoidal source on the simulation results: (a) $\textrm{ST}=T_0$, (b) $\textrm{ST}=10T_0$, (c) $\textrm{ST}=30T_0$, where $T_0$ is the period of the sinusoidal wave. The wave propagation direction is from left to right and the normalized field distributions are plotted in the $x$-$y$ plane ($z=2\lambda$, see Fig.~\ref{fig_domain}) and at the time step $t=1320\Delta t$.}
\label{fig_ST}
\end{figure}

The above presented results validate the developed parallel dispersive FDTD method and demonstrate the cloaking property of the structure. However, there are some numerical issues that need to be addressed in FDTD simulations. Besides the correction of numerical material parameters introduced earlier, since the cloak is a sensitive structure, for single-frequency simulations, the switching time of the sinusoidal source also has significant impact on the convergence time. Normalized field distributions from the simulations using different switching time are plotted at the time step $t=1320\Delta t$ and shown in Fig.~\ref{fig_ST}.
It can be seen that if the source is switched to its maximum amplitude within a short period of time, because of the multiple frequency components excited, and the cloak is essentially a narrowband structure due to its dispersive nature, the scattering from the cloak may occur, as shown in Fig.~\ref{fig_ST}(a). The scattered waves oscillate within the lossless cloak and hence it requires a very long time for the simulations to reach the steady-state. It is also demonstrated that if the switching time is greater than $10T_0$ where $T_0$ is the period of the sinusoidal wave, the scattered waves can be significantly reduced and a much shorter convergence time in simulations can be achieved. Therefore in the previous simulations, the switching time of $30T_0$ was used.

\begin{figure}[htbp]
\centering
\includegraphics[width=12cm]{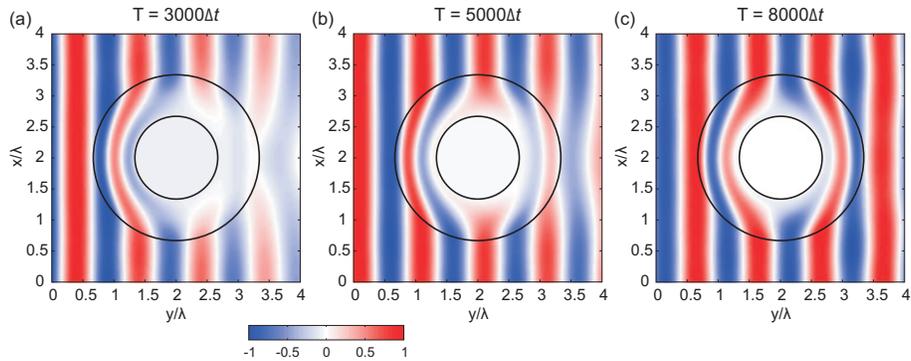}
\caption{Snapshots of the field distributions for the $E_z$ component at different time steps in the parallel dispersive FDTD simulations: (a) $t=3000\Delta t$ ($5.77$ ns), (b) $t=5000\Delta t$ ($9.62$ ns), (c) $t=8000\Delta t$ ($15.40$ ns), plotted in the $x$-$y$ plane ($z=2\lambda$, see Fig.~\ref{fig_domain}). The wave propagation direction is from left to right.}
\label{fig_time}
\end{figure}

Since the FDTD method is a time domain technique, it is convenient to study the transient response of the 3-D cloak. The snapshots of the field distributions for the $E_z$ component at different time steps $t=3000\Delta t$ ($5.77$ ns), $t=5000\Delta t$ ($9.62$ ns), $t=8000\Delta t$ ($15.40$ ns) are taken and plotted in Fig.~\ref{fig_time}.
It is shown in Fig.~\ref{fig_time}(a) that outside the shadow region behind the cloak ($y\sim3.5\lambda$, $x<0.5\lambda$ and $x>3.5\lambda$), waves propagate at the speed of light and the wave front remains the same as the one before reaching the cloak. However due to the fact that the waves that travel through the cloak undergo a longer path compared to the free space one, and since the group velocity cannot exceed the speed of light, the wave front experiences a considerable delay in forming back to the free space one in the shadow region behind the cloak, as it is illustrated by the field distributions at different time steps in FDTD simulations in Fig.~\ref{fig_time}. In fact, the convergence of simulations is quite slow and the steady-state is reached in simulations at around 13000 time steps ($25.02$ ns).

\begin{figure}[htbp]
\centering
\includegraphics[width=12.5cm]{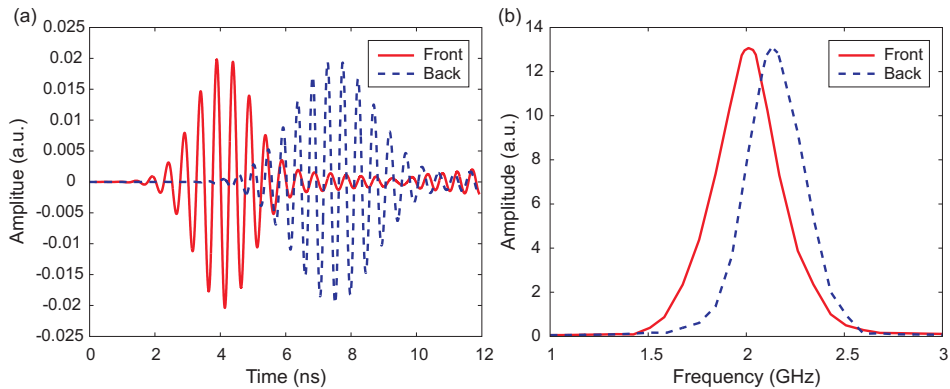}
\caption{(a) The recorded time domain signals at 5 FDTD cells away at both the front and back of the cloak along its central axis ($x=z=2\lambda$, see Fig.~\ref{fig_domain}). (b) The spectra of the recorded time domain signals.}
\label{fig_waveform}
\end{figure}

Another advantage of the FDTD method is that a wideband frequency response can be obtained with a single run of simulations. In comparison to the previous single-frequency case, a wideband Gaussian pulse with the central frequency of $2.0$ GHz and covering the frequency range of $1.5\sim2.5$ GHz is used instead. At 5 FDTD cells away at both the front and back of the cloak along its central axis ($x=z=2\lambda$, see Fig.~\ref{fig_domain}), the amplitude of $E_z$ is recorded during the simulation and then transformed to the frequency domain. The recorded time domain signals and their spectra are plotted in Fig.~\ref{fig_waveform}.
It is found from Fig.~\ref{fig_waveform}(a) that the time delay between the directly received time domain signal in front of the cloak (red solid line) and the received time domain signal through the cloak (blue dashed line) is approximately $3.4$ ns. However the distance between the two recording points is 0.402 m and the expected time delay for a plane wave propagating in the free space is $1.34$ ns. This clearly demonstrates that the 3-D cloak introduces a time delay to the waves propagating through it. It is also found from Fig.~\ref{fig_waveform}(a) that the width of the pulse passing through the cloak has been broadened, which is due to the dispersive nature of the cloaking material. It is interesting to note that in Fig.~\ref{fig_waveform}(b), the spectrum of the directly received time domain signal is centered at $2.0$ GHz, however the central frequency of the signal that passes through the cloak is considerably higher ($\sim2.1$ GHz). This frequency shift has been demonstrated theoretically in \cite{ZhangPRL}, which is explained as that the frequency components higher than the working frequency of the cloak are enhanced and the frequency components lower than the working frequency of the cloak are weakened. The shift is found to be much more pronounced from our analysis since the observation point is taken near the back surface of the cloak in our simulations, and it was $30\lambda$ away from the cloak considered in \cite{ZhangPRL}. The difference may be also due to the Lorentz dispersion model considered for the permeability of the cloak in \cite{ZhangPRL}, while in our simulations, both the permittivity and permeability are assumed to follow the same Drude dispersion model.

\section{Conclusion}
In conclusion, a parallel dispersive FDTD method has been developed to model the ideal 3-D cloak. The radial dependent permittivity and permeability of the cloak are mapped to the Drude dispersion model and taken into account in FDTD simulations using an ADE based method. Due to the memory restraint of a single PC, a parallel FDTD method is developed to handle the large amount of memory and simulation time required to model the 3-D cloak. FDTD simulation results are validated by those obtained using analytical methods. It is demonstrated that for single-frequency simulations, the source excitation needs to be switched on slowly enough to avoid the wave scattering from the cloak, due to the sensitivity of the cloaking material. It is also shown from the transient FDTD analysis that waves passing through the cloak experience a considerable time delay comparing with the free space propagations, as well as a pulse broadening effect and a blue-shift of the pulse's central frequency.

The ideal 3-D spherical cloak is considered in this paper. The method developed here can be also used to model cylindrical cloaks and compare their properties with the spherical one, such as the direction dependency issue. The ideal cloaks are lossless and the cloaking material properties vary in space continuously, which make their practical realization very difficult. The developed FDTD method can be also applied to study the effect of losses in cloaks, evaluate the performance of simplified cloaks as well as assist the design and realization of practical cloaks.


\end{document}